# Frequency domain measurements of thermal properties using 3ω-Scanning Thermal Microscope in a vacuum environment

G. Pernot [(1) *], A. Metjari [(1)], H. Chaynes [(1)], M. Weber [(1)], M. Isaiev [(1)] and D. Lacroix [(1)]

(1) Université de Lorraine, CNRS, LEMTA, F-54000 Nancy, France

Corresponding author *: gilles.pernot@univ-lorraine.fr

Abstract: Material thermal properties characterization at nanoscales remains a challenge even if progresses were done in developing specific characterization techniques like the Scanning Thermal Microscopy (SThM). In the present work, we propose a detailed procedure based on the combined use of a SThM probe characterization and its Finite Element Modelling (FEM) to recover in-operando 3ω measurements achieved under high vacuum. This approach is based on a two-step methodology: (i) a fine description of the probe's electrical and frequency behaviors in "out of contact" mode to determine intrinsic parameters of the SThM tip, (b) a minimization of the free parameter of our model, i.e. the contact thermal resistance, by comparing 3ω measurements to our simulations of the probe operating "in contact mode". Such an approach allows us to accurately measure thermal interface resistances of the probe as a function of the strength applied between the tip and the surface for three different materials (silicon, silica and gold). In addition, FEM modeling provides insights about the 3ω-SThM technique sensitivity, as a function of probe/sample interface resistance to measure material thermal conductivity, paving the way to quantitative SThM measurements.

# I. Introduction

The characterization of heat transfer and thermal properties at submicrometric scales[1–3] requires the development of specific devices with unprecedented temporal or spatial resolution. Time domain thermoreflectance technique using femtosecond lasers is now well-established [4–7] and studies of non-Fourier heat transfer flourish in the current literature [8–11]. On the other hand, scanning thermal microscopy (SThM) using local probes allows a spatial resolution solely limited by the size of its tip but still suffers of a lack of quantitative methods for analysis the outcome of the measurements. Designed with thermally sensitive elements (thermoresistive materials or thermoelectric junction), thermal probes can easily provide qualitative mapping of temperature contrast or thermal conductivity contrast of surfaces[12], biological medium[13], 1D materials and quantum dots [14]. Exhaustive review literature, about SThM techniques, are available [15–19] and provide an extensive description of the different experimental approaches, existing thermal probes and associated modeling[20–22]. Concerning the latter point, for years now, researchers have been trying to derive simple models to quantify the heat exchanged between a SThM probe and a sample. For instance, Lefevre et al [23] presented a one-dimensional thermal model representing a Wollaston thermoresistive probe constituted of Pt/Rh as thermally sensitive element. In order to derive a calibration expression relating the probe voltages and the thermal conductivity of the sample, the unknown parameters of the model (i.e. probe-sample contact radius and conductance) are determined by scanning various reference samples with known thermal conductivity. Even if such approach is well suited for this kind of tip, the major limitation of Wollaston's probes is the lateral resolution limited by the wire bend radius. Using a similar model, Puyoo et al [24] studied the thermal response of a Pd thermoresistive probe known as KNT (Kelvin Nanotechnology). This model was used in 3ω–SThM operating mode to determine the thermal conductivity of silicon nanowires considering the effect of the probe-nanowire contact size and the thermal interface resistance between the tip and the sample. Relying on an excellent lateral resolution, less than 100 nm [25], KNT have rapidly replaced Wollaston probes to perform thermal imaging. Nevertheless, its multilayer structure, composed of various materials and ultrathin layers (gold, NiCr, palladium and silicon dioxide) associated to a complex shape, limits the ability to develop simple theoretical models to describe the energy exchanged with a nanoscale sample. To overcome these issues, Ge et al [26] proposed a complex thermal resistance network in order to describe the probe-sample interactions and extract the thermal conductivity from the thermal spreading resistance but still with limitations. More recently, finite element method (FEM) with realistic 3-dimensional models were developed. One of the first report of such analysis has been proposed by Tovee et al [27]. In their study, they carried out a sensitivity analysis of the thermal signal obtained with KNT probe depending on the thermal conductivity of the sample in air and vacuum conditions. They even suggested to couple the probe with single-wall carbon nanotube to shift the sensitivity to higher thermal conductivity materials. Similar FEM models were used to quantify the effect of the cantilever thermal conductivity on the variation of electrical resistance of the probe [28] or the heating due to absorption of the AFM laser beam and the microscope stage temperature [29].

In this work, we propose to go further with a detailed study of SThM probe characterization associated with a Finite Element Modelling (FEM) under a high vacuum environment. This approach is based on a two-step methodology: (i) fine DC and AC calibrations of the probe behavior in "out of contact" mode to determine intrinsic parameters of the tip, (b) a minimization of the free parameter of our model, i.e. the contact thermal resistance, by comparing 3ω measurements to our simulations of the probe operating "in contact mode". Such approach allows us to determine accurately thermal interfacial resistance of the probe as a function of strength applied on the tip for three different substrates (silicon, silica and gold). The paper is organized as follow; after this introduction, a description of a KNT probe and its description for the FEM modelling is given. In this section, we describe the boundary conditions of the coupled heat transfer and electrical current and the main assumptions of the model. Simulated temperature maps, generated by Joule heating, are shown in both steady-state DC and transient AC modes. The third section is dedicated to the 3ω-SThM methodology description. There, we briefly present the experimental setup developed in this work and we derive the main mathematical equations of the 3ω-SThM technique. The fourth part details the two calibration steps that are used to characterize the main parameters of real probes. A new DC calibration is proposed to reproduce the behavior of a thermoresistive probe close to real experimental conditions. A calibration in AC mode is also proposed and we show that the frequency behavior of the KNT probe can be finely reproduce. The fifth part focused on the "contact mode" of the calibrated probe with a surface. In the latter, we first study the effect of thermal parameters such as thermal boundary resistance and thermal conductivity on the 3ω signal. Then, we investigate the frequency thermal response of silicon, silicon dioxide, gold and we evaluate the thermal interface resistance between the tip and these three kinds of materials. These thermal interface resistance measurements may provide a solid benchmark for further thermal property analysis of materials with a 3ω-scanning thermal microscope.

## II. KNT probe description and FEM modelling

Finite Element Method (FEM) with Comsol Multiphysics© is used to simulate the behavior of a SThM thermoresistive probe. The 3-dimensional model is constructed from information provided by Kelvin Nanotechnology®, MEB images of our probe under test and calibration methods which are detailed in the following sections. Briefly, the cantilever is made of a 400nm thick silicon dioxide layer. Note that in newer design, the cantilever is made in silicon nitride but the geometrical characteristics remain almost similar[30]. 140nm thick gold pads are deposited on top of a 5nm adhesion layer of NiCr. The thermoresistive part, located at the end of the pads, is composed of a 40nm thin layer of Pd that plays simultaneously the role of heater and sensor (see Figure 1).

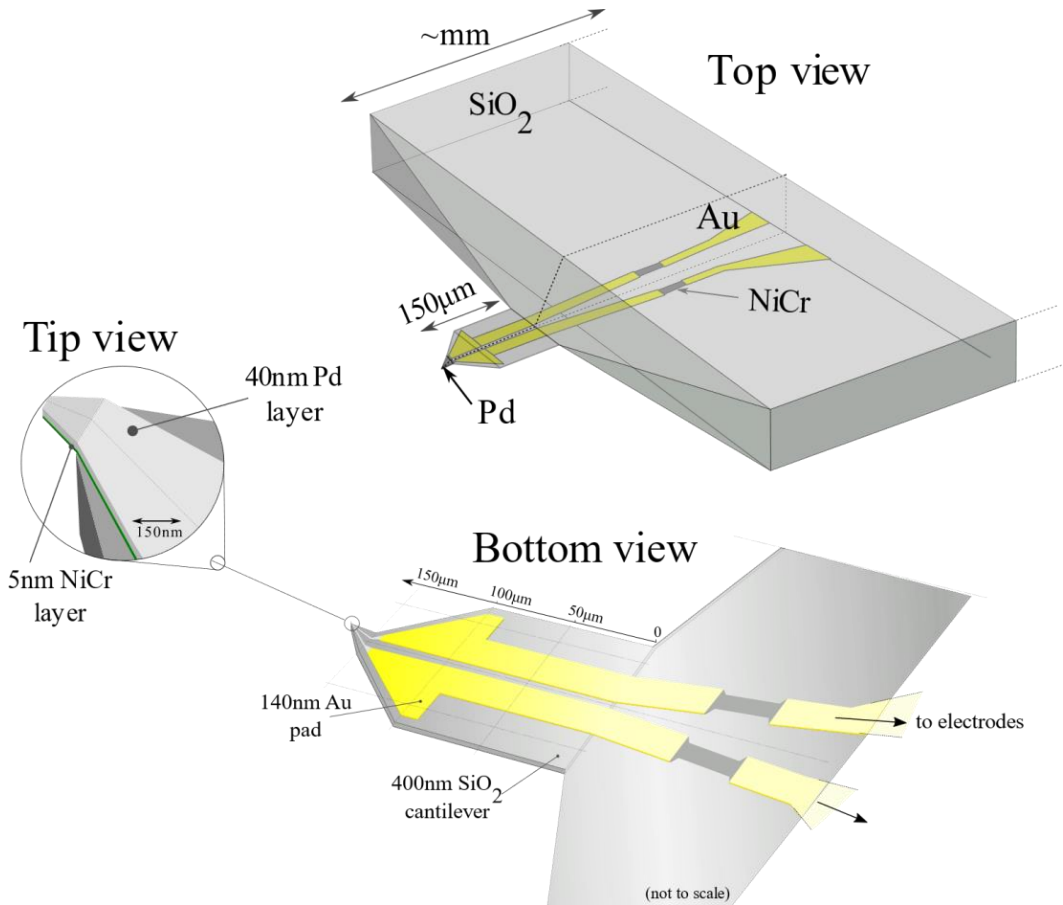

Figure 1: Schematic of a thermoresistive SThM probe.

The cantilever is connected to a silicon dioxide substrate and the gold lines are connected to two electrodes to allow a direct connection to a current or a voltage source. KNT probes also include two current limiters made in NiCr of about 100Ω each. Considering the entire probe, the resistance of the whole system approaches 300Ω.

A thin NiCr layer (~5nm), capped between the $SiO_2$ and the metallic Au and Pd parts, ensures their chemical adhesion on the substrate. A lack of quantitative information on transport properties of this layer remains. Using bulk properties, we estimate that its resistance approaches 10kΩ which means that only 3% of the total current flows into this layer, its effect on the electrical behavior of the probe is therefore negligible. To reduce the simulation time and increase the convergence of the FEM solver, this layer is not modelled explicitly but its influence is embedded in the electrical characterization of the probe. Indeed, the identified temperature coefficients (TEC) of the electrical resistance "capture" the presence of this layer and its effects on the thermal response of the probe.

The major advantage of FEM is the ability to solve complex multiphysic problems. In our case, our model couples the equations for heat transfer and electrical current to simulate Joule heating. The different electrical and thermal boundary conditions are given on Figure 2. Using the symmetry of the KNT probe, only half of system should be considered and the net heat flux is set to zero along this symmetry plane. At the end of the simulated geometry, the temperature is set to be constant and equals to the surrounding temperature. Under vacuum condition, free convection is neglected so that adiabatic boundaries can be safely considered on all the free surfaces. We verified this assumption experimentally by measuring the SThM signal for various pressures from atmospheric condition (1015mbar) to high vacuum ($5 \times 10^{-5}$mbar). As reported by Nakabeppu and Suzuki [31], we found that the signal remains constant below 0.1 mbar proving that the influence of free convection becomes negligible below this pressure.

Radiative effects are also not considered here; this is supported by the low emissivity of the constituting materials (metals and silicon dioxide). Radiation exchanges with the surrounding can also be neglected as the exchange surface view factors are very small. Additionally, as for the effect of the NiCr layer, radiative losses are "embedded" in the behavior of the probe during the calibration procedure of the TEC. For what concern electrical boundary conditions, we imposed the incoming current, while at the other end of the geometry, we set a ground voltage. We checked that numerical modelling conserves the current at the output of the geometry during the simulation procedure.

The electrical and thermal properties for Au, SiO$_2$ and NiCr limiter have been kept to the Comsol® database values. On the contrary, for the Pd line, the electrical conductivity has been modified according to Eq. (2) and the temperature coefficients α and β have been adjusted using a new DC calibration method explained in the following section.

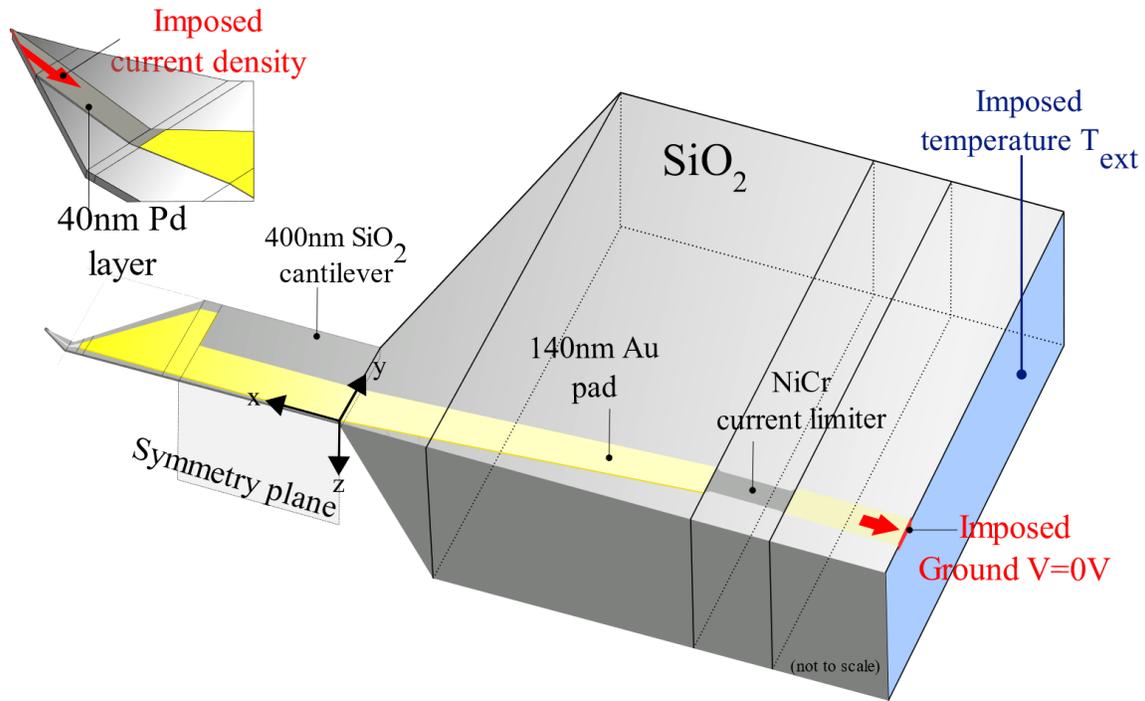

Figure 2: schematic representation of the 3d geometry built for fem analysis and the boundary conditions applied for thermal and electrical modeling.

Before dealing with 3ω measurements and time-dependent simulations, the FEM model of the probe was tested in steady-state mode to evaluate typical temperature elevations and hotspots as a function of the current input. Figure 3 (left) shows the calculated temperature map arising in the structure with a typical current of $I_0 = 1mA$. As expected, the simulation shows that the heated region is located at the apex of the probe while the temperature rapidly decreases along the gold line.

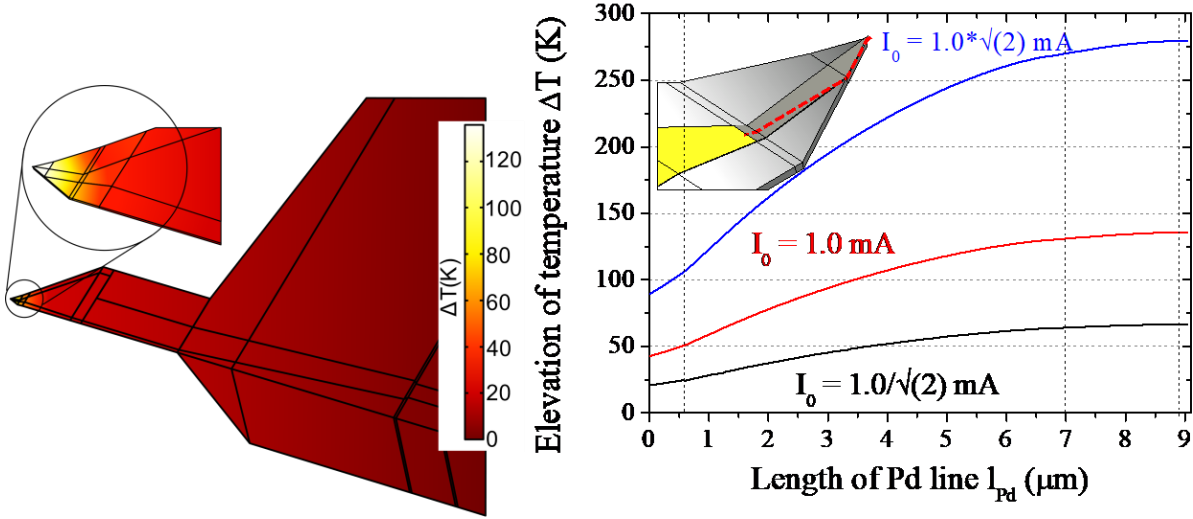

Figure 3: (left) 3d temperature map calculated in the probe. (right) Elevation of the temperature along the Pd layer (see inset) for 3 different currents.

The temperature profile along the Pd layer, calculated for 3 different currents, is showed on Figure 3 (right). One can see that very large temperature gradients take place over only few micrometers. For a current of about 1.4mA, the elevation of temperature at the very end of the tip (i.e. apex) reaches 280K while the difference of temperature between each Pd ends is about 200K. Such temperature gradients demonstrate that currents above 2mA can highly damage the probe. Even for 1.5mA current, we have witnessed strong changes in the resistance during a daily use. Note that those changes mostly affect the value $R_0$ while the probe's TEC were not affected. In our case, we have chosen to limit the current values below 1mA. Under this condition, no significative changes in the electrical properties of the probe have been experienced for weeks. Interestingly, for such current, while one can expect a hotspot at the apex of about 200K above the ambient temperature, the variations of the total resistance (see Figure 5), measured on a real probe, are only about 5Ω (~2% of variation for a typical 300Ω total KNT resistance). For this reason, AC current and third harmonic voltage measurements, known as 3ω-SThM, usually are preferred.

## III. 3ω-SThM: method and model

The core of our setup is developed from a XE-100 Park Systems model that has been customized to fit in a Pfeiffer high vacuum chamber. The 3ω technique can be easily applied by injecting an AC-current in the thermoresistive probe with the cyclic frequency ω. The signal is generated by a Stanford Research System DS360 chosen for its very low distortion rate. As for classical 3ω setup, the probe is inserted into a Wheatstone bridge to filter the 1ω component. In our case, the bridge has been developed in the lab and the resistances have been carefully chosen for their very low temperature coefficient (typ. ±5ppm/°C) to avoid 3ω parasitic signals due to the resistance's self-heating. Then, the signal coming out of the bridge is amplified (x100) and finally feed to a lock-in amplifier (Signal Recovery 7280) which measures the amplitude and phase of the 3ω voltage.

We, now, discuss of the fundamental aspects of the 3ω technique when applied to a SThM thermoresistive probe[32]. Let's consider an AC-current of the form $I(t) = I_0 \cdot \cos(\omega t)$, flowing in the probe, $I_0$ is the amplitude and ω is the pulsation of the first harmonic. Due to Joule heating, the temperature field inside the probe at coordinate $x_0$ oscillates at a pulsation 2ω and can be written as:

$$\Delta T(x_0, t) = T_{rms}(x_0) + T_{2\omega}(x_0) \cdot \cos(2\omega t + \varphi_{2\omega}(x_0)) \quad (1)$$

Where ΔT is the difference between the temperature at coordinates $x_0$ and the environment, $T_{rms}$ is the average elevation of temperature at $x_0$ and $T_{2\omega}$ is the amplitude of the oscillation at pulsation 2ω. Note that considering the thickness, width and the direction of current flow, only a 1D temperature field along the Pd line can be considered. Temperature profiles, calculated with FEM along the other directions, have verified this assumption.

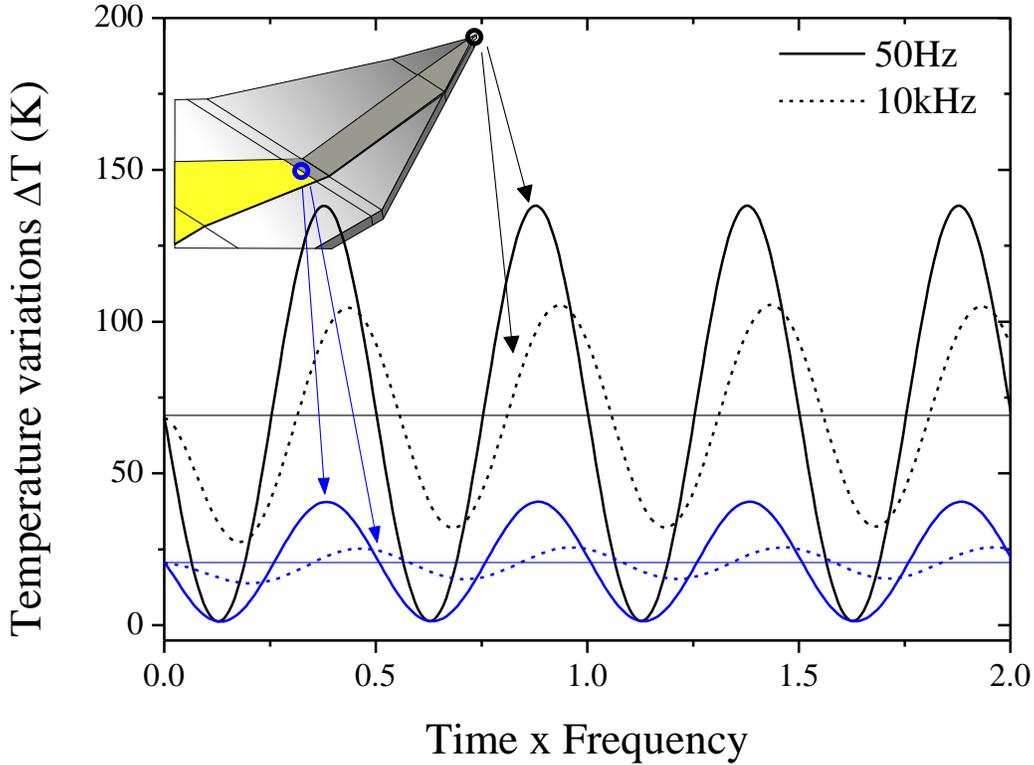

Figure 4: Temperature oscillations at the extremity of the Pd line calculated for low frequency and high frequency.

Figure 4 shows elevations of temperature $\Delta T(x_0)$ calculated, with the previously discussed FEM model, at both end of the Pd line (black and blue dots) for an amplitude of the current of 1mA and for 2 different excitation frequencies (low 50Hz and high 10KHz). Those values have been chosen below and above the probe's cut-off frequency and demonstrate the drop of the 3ω voltage at high frequencies (see Figure 6). It is worth noticing that the "rms" values of temperature $T_{rms}$ (straight lines on Figure 4) are independent of the frequency and can be obtained by injecting a DC-current $I_{rms}=I_0/\sqrt{(2)}$. This property will be useful for increasing the convergence of transient simulations, since it provides an excellent initialization of the variables by eliminating the "transient" response that normally occurs with a random initialization of the variables.

Then, for the temperature dependence of electrical resistivity of the Pd thin film, we assume a polynomial form which can be expressed as:

$$\rho(T) = \rho_0[1 + \alpha(T - 273.15K) + \beta(T - 273.15K)^2] \qquad (2)$$

Where $\rho_0$ is the electrical resistivity at 273.15K and T is the temperature in K, α and β are respectively first and second order temperature coefficients (TEC). Considering the complex shape, an analytical -temperature dependent- expression for the resistance cannot be easily obtained by integration of the latter expression. But, according to Eq (1), a general expression for *R(T)* can be expressed:

$$R(t) = R_0 + R_{rms} + R_{2\omega} \cdot \cos(2\omega t + \phi_{2\omega}) + \Theta(4\omega) \qquad (3)$$

Where $R_0$ is the resistance at the temperature of the environment (i.e. zero current). $R_{rms}$ is the variation of resistance due to the average elevation of temperature in the probe $T_{rms}(x_0)$ and $R_{2\omega}$ is the amplitude of oscillation of the resistance resulting of the temperature field at 2ω. The term $\Theta(4\omega)$ stands for the fourth harmonic contribution which is often neglected.

Finally, in 3ω experiment, the measured quantity is the voltage across the probe. From Ohm's law, it is straightforward to see that the voltage is the sum of the voltage oscillating at the fundamental harmonic $V_\omega(t)$ and a component oscillating at the third harmonic $V_{3\omega}(t)$. Using the definition of the resistance given in equation (3), each voltage component can be expressed as:

$$V_\omega(t) = (R_0 + R_{rms})I_0 \cdot \cos(\omega t) + \frac{1}{2}R_{2\omega}I_0 \cdot \cos(\omega t + \phi_{2\omega}) \qquad (4)$$

While the third harmonic component depends only of the amplitude oscillation of the resistance $R_{2\omega}$:

$$V_{3\omega}(t) = \frac{1}{2}R_{2\omega}I_0 \cdot \cos(3\omega t + \phi_{2\omega}) \qquad (5)$$

Due to the higher order of temperature dependence of the electrical conductivity, fifth and higher harmonic exist but their amplitude is negligible.

# IV. Out-of-contact studies - DC and AC calibration methodology

The experimental device and the simulation tool presented in the previous section are not self-sufficient to allow the evaluation of material thermal properties from 3ω-SThM measurements. Each probe used for such purpose needs a calibration procedure because it exhibits a different electrical behavior that influence the thermal response and the recorded 3ω signals. In this part, we describe the calibration in DC mode from which we estimate the temperature coefficients and in AC mode from which we can study the frequency behavior of a KNT probe.

### 1. DC calibration

The most straightforward method to determine the TEC is to calibrate the probe in a temperature chamber and measure its changes of resistance versus temperature[25]. One drawback is that the entire probe is heated and consequently the TEC of NiCr and gold must be characterized separately. In addition, temperature chambers work under an ambient environment which induces unknown convection losses that do affect the resistance of the probe. Usual values of TEC for Pd and NiCr, measured using a temperature chamber, can be found in the literature to $1.2 \times 10^{-3} K^{-1}$ and $0.15 \times 10^{-3} K^{-1}$ respectively[33].

We choose a slightly different approach where the TEC are extracted under similar conditions as during SThM experiments. First, we use a micro-probe setup, with a 10µm radius tip, to measure separately the resistance of each part of the probe under a current set to 100µA. With this current, the self-heating of the probe can be neglected [34]. From this step, the resistance of the Pd line and the resistance of each current limiter can be estimated with an uncertainty of about ±5Ω. This uncertainty corresponds to a standard deviation value over 3 or more different measurements performed on each part.

Then, in the vacuum chamber, we vary the DC current while measuring the changes of resistance of the probe. During this calibration, the probe is kept out-of-contact i.e. far from any surface that may influence the measurement. We vary the current from 200µA to 1mA and monitor the resistance using a Keithley A2400 source-meter set in 4-wire mode. We use a least-square fitting algorithm coupling Matlab® and Comsol® with 2 free parameters, α and β from equation (2), to find the best fit (see Figure 5). TEC values are found α=0.624x10$^{-3}$K$^{-1}$ for the first order and β=-6.1x10$^{-7}$K$^{-2}$ for the second order TEC. Similar results for TEC have been obtained on other probes from the same batch but not used in this work. Table 1 summarized the calibration results on the probe we used in this work, and another one from the same batch to highlight the similarities of the measured resistances and identified TEC.

|  | Total resistance (Ω) | Cantilever resistance (Ω) | Limiter resistance (Ω) | 1$^{st}$ order TEC for Pd α (K$^{-1}$) | 2$^{nd}$ order TEC for Pd β (K$^{-2}$) |
|---|---|---|---|---|---|
| Probe used in this work | 299 | 106.0±5.0 | 92.5±5.0 | 0.624x10$^{-3}$ | -6.1x10$^{-7}$ |
| Other probe (not used in this work) | 310 | 117.0±5.0 | 92.5±5.0 | 0.620x10$^{-3}$ | -3.35x10$^{-7}$ |

Table 1: Electrical properties of SThM probe identified in this work. The influence of ±5% variation on the 1$^{st}$ order TEC coefficient is shown on Figure 5.

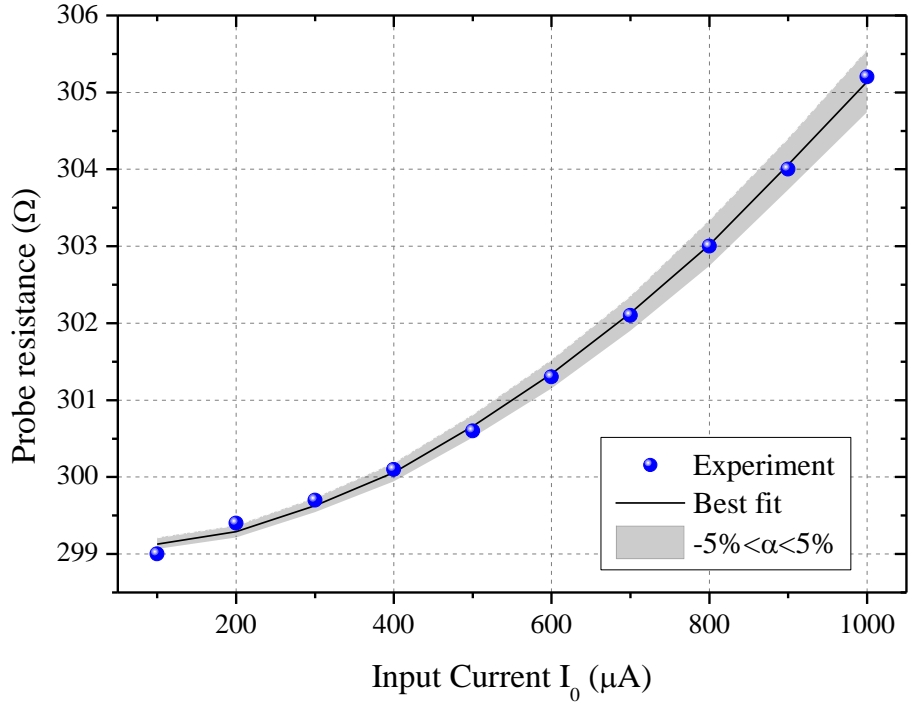

Figure 5: DC calibration of the probe and best TEC parameters evaluated for Pd. The parameters for the best fit are given in Table 1.

It is worth noticing that our TEC values for Pd differs from a factor two compare to those reported earlier. As stated before, this difference can be explained by the choice of experimental approach; while our method got rid of the convection losses and minimizes the influence of the TEC of the other parts of the probe, temperature chamber's measurements can be biased by those unknown parameters. On the other hand, our calibration method does not give the absolute TEC value since the temperature distribution in the Pd heater is not homogeneous and therefore the usual TEC definition of a resistor $\frac{1}{R_0} \cdot \frac{\Delta R(T)}{\Delta T}$ cannot be used in this case.

2. AC calibration

In AC current, we studied the frequency behavior of the probe by varying the frequency from 50Hz up to 10kHz with a constant current amplitude of 1mA. The results are shown on Figure 6 and the 3ω voltage is plotted versus frequency. The frequency domain response exhibits a low-pass filter behavior with a cut-off at 3.6 kHz.

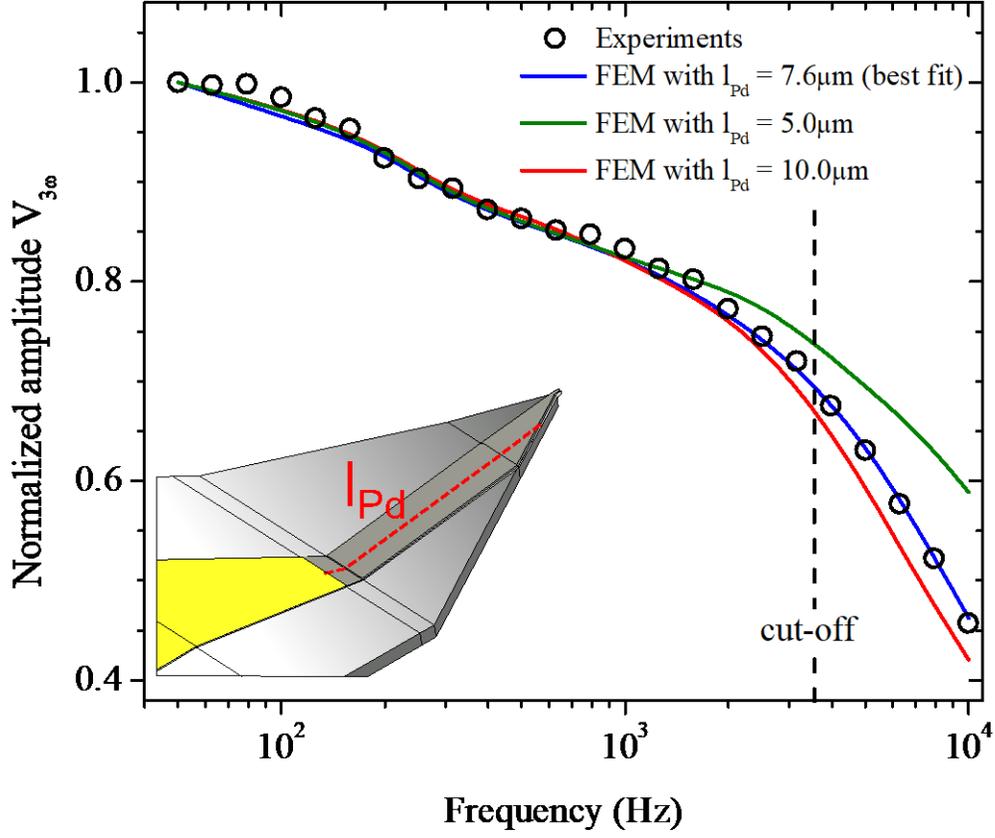

Figure 6: Normalized 3ω amplitude signal in ac-mode for various frequencies. The black dots show the experimental signal obtained in vacuum. The red triangles represent the initial set of simulation for a Pd length $l_{Pd}$, as shown in the inset, of 10µm. The blue dots indicate the best fit obtained for a Pd length of 7.6 um

From the model, the evaluation of the 3ω amplitude is not straightforward since only the total potential ($V_\omega(t)+V_{3\omega}(t)$) is evaluated by the finite element analysis and $V_{3\omega}$ is known to be about 100 times weaker than $V_\omega$. In addition, we showed in the previous section that the link between the temperature field, the oscillation of the resistance $R_{2\omega}$ and the 3ω voltage is not straightforward. To overcome those problems and evaluate the 3ω signal, we ran transient simulations over 3 frequency periods using a temporal resolution of 120 time-steps per period. With this resolution, the temperature field and the 3ω voltage are sufficiently defined with respectively 60 points and 40 points per period. To avoid numerical interpolation of the variables, we set a strict time-stepping in the transient solver used by Comsol®. Finally, as explained before, to increase the convergence of the solver, the variables are initialized using a steady-state calculation with an input DC current equal to $I_{rms}$. Once the simulation is done, we post-process the results using a Levenberg-Marquardt minimization algorithm in Matlab® to extract the amplitude and phase of the 3ω voltage out of the calculated total voltage. A similar procedure was also applied to study the influence of the NiCr limiter on the 3ω signals. For a current limiter of about 100Ω with a TEC of $1.4 \times 10^{-4} K^{-1}$ and current amplitude of 1mA, we have found that the typical $V_{3\omega}$ amplitude, due to the limiter, is about 10 times smaller than the 3ω voltage generated by the Pd film.

By repeating this procedure for all frequencies from 50Hz up to 10kHz, we show that the model can reproduce the low-pass filter behavior of a KNT probe. For this calibration step, the main goal was to adjust the cut-off frequency. The influence of current intensity, resistance or TEC on the frequency response have been studied, and it was found that they mostly influence the amplitude of the 3ω voltage.

Like reported by Lefevre et al. for a Wollaston probe [21], the variations of $V_{3\omega}$ of a KNT probe with frequency and its cut-off are sensitive to the heating area; i.e. geometrical parameters of the Pd line (thickness, width and length). We chose to keep the thickness constant and equal to 40nm as given by the manufacturer. The width was also kept constant and set to 1.55um, this value has been averaged from MEB images performed on previous probes and found to be almost unchanged from one probe to another. Therefore, we chose to vary the length of the line $l_{Pd}$. Our choice is supported by the fact that this parameter can hardly be measured from MEB images since, at the Au/Pd junction, a part of the Pd has been covered by Au during the deposition while the transition between Au and Pd is well located in the FEM model. The results for 3 values of $l_{Pd}$ (5µm, 7.6µm and 10µm) are shown and compared to the experimental $V_{3\omega}$ results on Figure 6. The best fit was obtained for a value of $l_{Pd}$ of 7.6 µm, it shows a remarkable agreement with the experimental data obtained in vacuum.

# V. Study of contact mode and thermal interface resistance measurements

At this stage, the calibration steps are completed and the main parameters that characterize a KNT probe's response have been identified. We carried out measurements on various materials, with known thermal properties, to investigate the contact between the probe and the surface. As for the calibration steps, these studies were conducted under a high vacuum environment where free convection is neglected and the water meniscus, that appears under ambient conditions, is removed so that the supplementary solid/liquid channel for the heat transfer can also be neglected.

Unlike usual SThM studies, we did not perform raster scans to produce thermal images. Instead, we carried out 3ω measurements on a single point of the surface allowing only a vertical displacement to control the contact force. It is assumed that the signal, when the contact between the probe and the sample is achieved, depends on the thermal properties of the sample but is also strongly dependent on the contact interface resistance between the probe and the sample. These effects are numerically studied by varying the contact thermal resistance in the finite element model and presented on Figure 7.

The modelling is achieved by coupling the probe's model to a solid block representing a sample with different thermal conductivity in the range of 0.1 to 1000 $W.m^{-1}.K^{-1}$. The dimensions of the sample are large enough so that it may be considered as a heat sink. The thermal interface resistance is set on the contact area between the probe and the surface of the sample. The results are plotted for a single frequency taken at 50 Hz and the 3ω voltages are normalized with the out-of-contact value calculated at the same frequency. These results are in agreement with those presented by Tovee et al [35] who reported a drop of the probe's temperature as the sample thermal conductivity increases. A similar behavior is observed for the 3ω voltages and establishes the range of sensitivity of a KNT probe regarding the thermal conductivity of the sample under study.

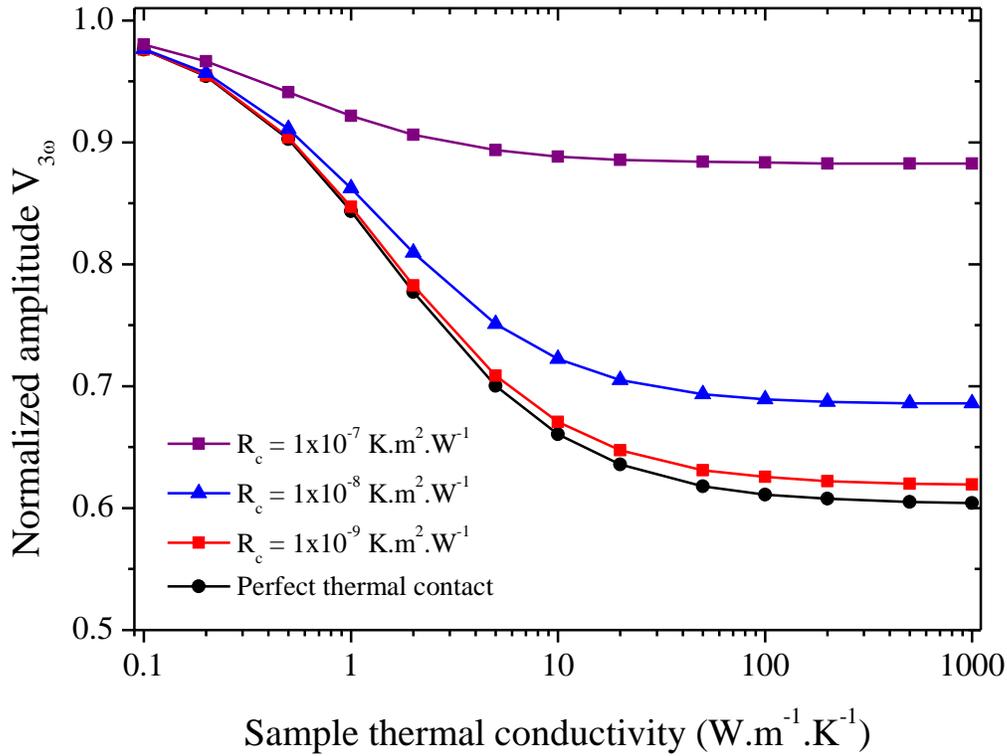

Figure 7: 3ω amplitude signals versus sample thermal conductivity for 4 thermal interface values. The data are calculated for a frequency equals to 50Hz and are normalized by the out-of-contact signal.

Figure 7 also demonstrated the critical influence of the thermal interface resistance between the tip and sample. We varied the interface resistance value Rc from 0 (i.e. perfect thermal contact) up to $10^{-7}$ K.m$^2$.W$^{-1}$. For a perfect contact, one can expect a variation of the 3ω signal of about 40 % between the out-of-contact and contact signals for the most conductive materials. For interface resistances around $10^{-9}$ K.m$^2$.W$^{-1}$, the same order of variation of the signal is found. The latter order of magnitude for Rc is similar compared to those measured by TDTR between a metallic transducer (typically Al) and a semiconductor layer [9]. For higher values of the interface resistance, the voltage exhibits a significant drop and the sensitivity of the SThM measurement is highly reduced. For example, at $10^{-7}$ K.m$^2$.W$^{-1}$ only 10% difference in the measured signals can be expected between highly conductive materials and thermal insulators. This plot also highlights why the SThM technique has been mostly utilized for thermal property characterization of low thermal conductivity (< 1W.m$^{-1}$.K$^{-1}$) materials [12] since, due to a higher thermal resistance, the impact of the interface on the signal is lowered. Therefore, reducing thermal contact resistance between the tip and the sample is a major issue to perform precise measurements of the thermal conductivity with a thermal microscope.

We conducted SThM experiments on three materials with a wide range of thermal conductivity and different nature: silicon and silicon dioxide substrates and a 140nm thin film of gold. Our goal was to evaluate the probe/surface thermal interface resistance. The data presented hereafter are obtained by maximizing the contact, i.e. minimizing the interface resistance, between the tip and the surface by following the approach procedure described on Figure 8.

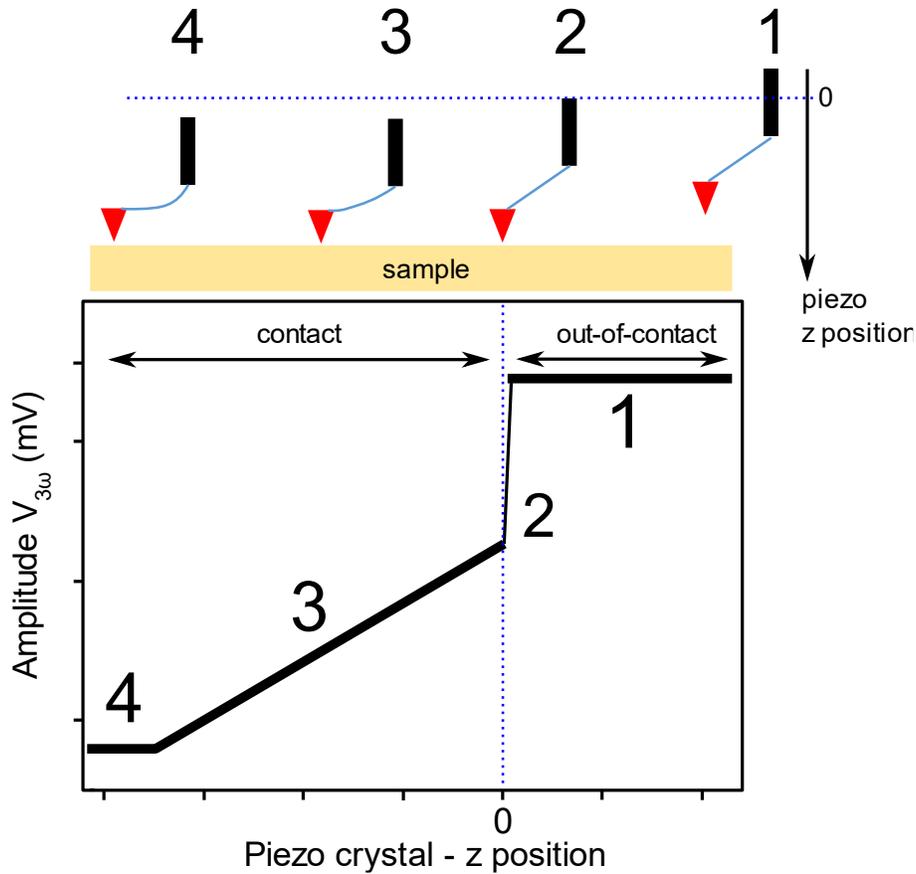

Figure 8: Manual tip to sample approach and minimization of the contact in vacuum.

The contact method exhibits 4 significative steps as the piezoelectric crystal, that controls the probe z-position, is moved. First, when the probe is far from the surface, the 3ω voltage is constant and maximum. At step 2, the signal exhibits a sudden drop which corresponds to the contact between the tip and the surface (usual SThM thermal images are produced at this step). During step 3, we kept moving the piezoelectric crystal after the contact resulting in a bending of the cantilever but also a quasi-linear decrease of the signal which corresponds to a reduction of the interface resistance. Finally, the signal reaches a minimum (step 4) and constant value corresponding to the minimum of thermal interface resistance that can be achieved. Note that this minimum signal is repeatable over several days of experiments. An example of the minimum contact signal obtained, with a 1mA current amplitude, on a silicon substrate, and compared to its out-of-contact signal, is shown on Figure 9.

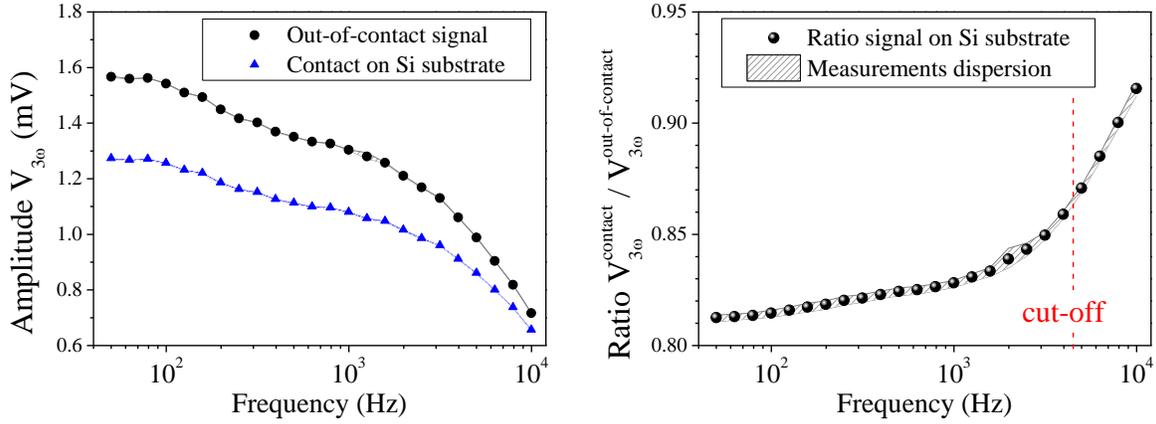

Figure 9: (Left) single point measurement of the 3ω voltages obtained out-of-contact and on a Si substrate using the minimization of the contact for a current of 1mA. (Right) ratio between contact and out-of-contact signals. The ratio has been averaged on 3 different spots and the shaded area corresponds to the dispersion of the measurements.

Measurements were carried out on 3 different spots on the surface using the same contact minimization procedure. The graph on the left shows the average contact signal (blue triangles) and the out-of-contact curve (black dots) obtained by moving the probe few micrometers away from the surface. The shape of the thermal response over the frequencies is very similar to the out-of-contact signal, only the amplitude is lower due to the heat flux transferred to the sample. These two curves demonstrated that the shape of the thermal response is mostly dominated by the probe. In order to suppress the effect of the "probe" and simplify the analysis of the measurement, it is convenient to examine the ratio between contact and out-of-contact of the 3ω signals as shown on the right graph on Figure 9.

The ratio $V_{contact}/V_{out-of-contact}$ is almost independent of the current and weakens low-pass filter shape of the signal. Only at high frequencies, the ratio increases toward 1. This effect can be understood by the fact that at high frequencies, the 3ω amplitude is strongly reduced by the low-pass behavior and therefore the probe becomes less dependent of its surroundings. It is also important to note that the ratio signal representation overcomes the lack of repeatability of SThM measurements as the 3ω voltage often varies from one day to another with the changes of the experimental conditions such as temperature's chamber, fluctuation of current, etc.

Assuming known thermal properties of the material, it is now possible to evaluate the interface resistance by comparing the ratio signal with the one calculated using Comsol®. The fitting procedure is carried out by varying manually the value of the interface resistance until a good agreement with the experimental curve is found. Considering the high sensitivity of the signal to the contact resistance, only few trial runs are necessary to get to the optimum value. The left graph on Figure 10 shows the adjusted curve on Si sample. The best fit is obtained for a resistance $R_c = 4.5 \times 10^{-8} K.m^2.W^{-1}$. Very few literature exists on this topic [36,37], the metal/semiconductor interface resistance is about 10 times higher than the values measured by TDTR (typ. $5 \times 10^{-9} K.m^2.W^{-1}$) between the same type of materials[8]. Considering the fact that for TDTR samples, the metallic layer is usually deposited by evaporation technique, the solid/solid contact is almost perfect and thus, the interface resistance is solely due the lattice mismatch that exists between the two materials. We also show the sensitivity of the interface resistance on the experimental signal by varying the best-fit value by ±10%. The shaded area shows the effect of such variations on the ratio signal and the dispersion of the measurements are included into this ±10% "error bars". Only high frequencies above the cut-off exhibits a larger discrepancy with the calculated ratio and deviates from the estimation range. This discrepancy may be explained by tiny errors in the parameters of our model (geometry, electrical boundary resistances…) which impacts the high frequencies. But considering low-pass filter and the reduction of the sensitivity of a KNT probe at high frequencies, it is reasonable to admit that this discrepancy does not alter the validity of our fits and of our identified thermal interface resistances.

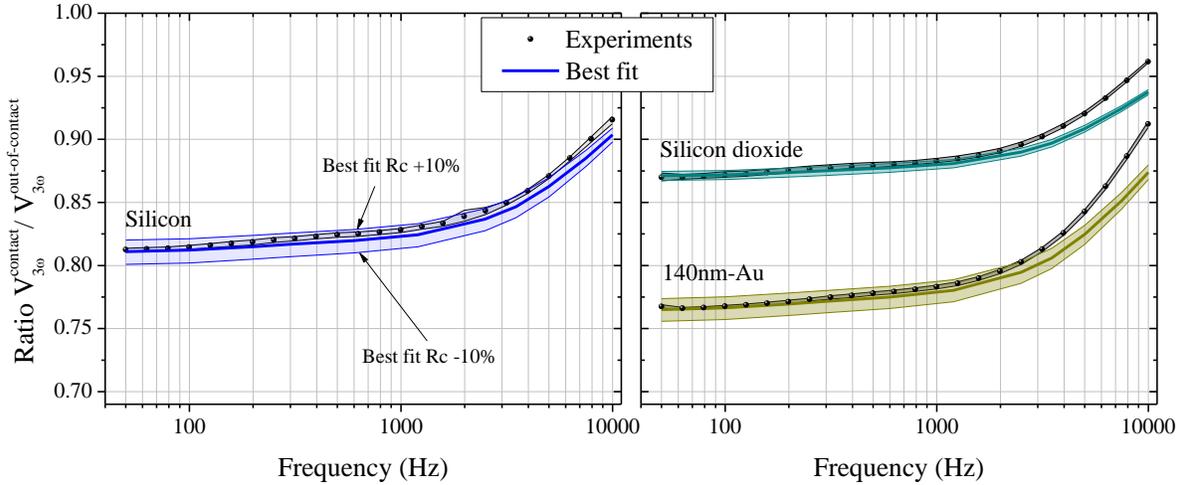

Figure 10: (Left) experimental ratio on silicon (black dots) and best fit (blue line). The blue area represents the ratio calculated using Rc±10% (Right) Results and best fit obtained on silicon dioxide and on 140nm thin gold layer (the results have been separated for the sake of clarity but the y-scale is similar on both graph).

Similar experiments and analysis were conducted on an amorphous silicon dioxide. In this case, the minimum interface resistance, which can be achieved, is $3.5 \times 10^{-8} K.m^2.W^{-1}$. This value is quite similar with the one, we found, for the silicon substrate. The last sample studied in this paper originates from a broken KNT probe and measurements were conducted on the gold pad. As expected, the metal/metal interface resistance is found to be lower than the two other materials, Rc = $2.75 \times 10^{-8} K.m^2.W^{-1}$, as the thermal energy can be more easily transported by the free electrons from one metal to the other. Table 2 summarized the minimum interface resistances that have been identified in this work.

|  | Rc thermal interface resistance ($K.m^2.W^{-1}$) | | |
|---|---|---|---|
|  | Si substrate | $SiO_2$ substrate | 140nm Au layer |
| Step 2 – tip to surface minimum contact | $18 \times 10^{-8}$ | $50 \times 10^{-8}$ | $11 \times 10^{-8}$ |
| Step 4 – tip to surface maximum contact | $4.5 \times 10^{-8}$ | $3.5 \times 10^{-8}$ | $2.75 \times 10^{-8}$ |

Table 2: Summary of the identified thermal interface resistance for 3 different materials and given for minimum contact (step 2) and maximum contact (step 4).

As this minimization approach method can hardly be used to perform raster scans and thermal imaging, we also give an estimation of the interface resistances that may be measured, at step 2 on Figure 8, when the tip just reaches the surface. In all cases, the "contact" interface resistance is found to be higher than $10 \times 10^{-8}$ K.m$^2$.W$^{-1}$. We would like to point out that these values are only indicative since in most AFM control software, the force between the probe and the surface is an setting that can be adjusted by the user and therefore the interface resistance may change from one setting to another. Nevertheless, the range of interface resistance values given here highlights the reduction of sensitivity of the SThM and points out the difficulties of quantitative analysis of thermal microscopy. Under ambient conditions, the interface resistances may decrease due to the influence of a water meniscus which increase the contact area and create a new channel solid/fluid/solid for the heat to be transferred to the sample[38–40].

## VI. Conclusion

In this study, we have achieved a detailed study of Scanning Thermal Microscopy in vacuum from both numerical and experimental points of view in order to predict accurately the tip/sample interface thermal resistance. This work is focused on standard commercial SThM thermoresistive probes known as "KNT" but the methodology can easily be applied to other kind of SThM probes. The study under a vacuum environment freed our analysis from convection losses and water meniscus occurrence that remain poorly understood. These specific issues will be addressed in future works.

First, we have described the finite element model build to accurately mimic the thermal behavior of KNT probes. The 3-dimensional multiphysic model, developed with Comsol®, couples heat transfer and electric current equations to simulate the Joule heating that takes place in the thermoresistive probes. Numerical simulations, using both DC and AC currents, have demonstrated that: (i) very large temperature gradients are generated in the active Pd film explaining sudden changes in properties that often occurs when using this kind of probes, and (ii) due to a limited temporal thermal response of the system, the temperature field oscillating at the second harmonic is strongly reduced at frequencies above the cut-off frequency. The latter feature explains the low-pass filter behavior already reported for 3ω-SThM with KNT probes.

In the second part of this study, we have coupled SThM experiments performed under high vacuum and FEM simulations to calibrate the main electrical and thermal parameters of the probe. We describe two original calibration steps. By monitoring the variations of the electrical resistance with DC currents, we have evaluated the two temperature coefficients (first and second order) of the electrical conductivity of the thin Pd layer. TEC values reported in this work, even though repeatable from one probe to another, are found to be by a factor of 2 smaller than those reported in earlier literature. We pointed out that our calibration method is different from others because it does not allow us to measure the temperature coefficient usually define for a resistor set at a uniform temperature. The second calibration step is performed in 3ω mode. In this stage, the cut-off frequency of the probe is adjusted by slightly modifying the length of the Pd leg. It is worth noticing that for both calibration steps, remarkable agreements are obtained between real SThM measurements and fitted numerical results.

Finally, we have applied our methodology to study the contact effect between a KNT tip and the surface of various known materials. Again, those experiments have been carried out in vacuum to overcome the effect of water capillarity between the tip and sample which strongly affects heat transfer. In this part, we have demonstrated that the 3ω voltage obtained in contact in frequency domain is convoluted with the probe frequency response. To simplify the analysis, we showed that the ratio between contact and out-of-contact signals is a useful quantity to provide quantitative analysis of the heat transferred from the tip to the sample. Fitting this ratio in the frequency domain to compare experiment and FEM simulation, we were able to provide reliable measurements of the tip/surface contact resistance for silicon, silicon dioxide and gold. For usual contacts, the range of interface resistance is found to be extremely high ($> 10^{-7}$ $K.m^2.W^{-1}$) and therefore explains why the KNTs have been mostly used to study polymers and other low thermal conductivity materials. We have also showed that it was possible to manually reduce the thermal interface resistance up to a minimum value and therefore increase the sensitivity of the SThM technique. The resistances in that case were found to be one order of magnitude lower ($> 10^{-8}$ $K.m^2.W^{-1}$) and in the same range for all three materials studied here. We believe that the methodology and results presented in this work may provide a solid benchmark for future SThM thermal property analysis.

# VII. Acknowledgements

The authors want to thank Dr. Nicolas Stein and Dr. Stéphane Grauby for the insightful discussions and advices. We also would like to thank Jamal Ouhajjou for his help in the development of the SThM setup. This work is supported by FEDER-FSE Lorraine et Massif des Vosges 2014-2020 and the French National Research Agency and the Swiss National Science Foundation through the ANR project "3D-Thermonano" (grants ANR-17-CE05-0027 and 200021E-175703/1) and the ANR project SPiDER-man (grants ANR-18-CE42-0006).

# VIII. Data availability

The data that support the findings of this study are available from the corresponding author upon reasonable request.